\documentclass[twocolumn,superscriptaddress]{revtex4-2}
\usepackage{epsfig}
\usepackage{amsmath}
\usepackage{amssymb}
\usepackage{graphicx}
\usepackage{dcolumn}
\usepackage{bm}
\usepackage[usenames]{color}
\usepackage[breaklinks]{hyperref}
\hypersetup{colorlinks=true,allcolors=blue}
 
\begin{document}
	
\title{High-Chern number phase in the topological insulator multilayer structures}

\author{Yi-Xiang Wang}
\email{wangyixiang@jiangnan.edu.cn}
\affiliation{School of Science, Jiangnan University, Wuxi 214122, China}
\affiliation{School of Physics and Electronics, Hunan University, Changsha 410082, China}

\author{Fuxiang Li}
\email{fuxiangli@hnu.edu.cn}
\affiliation{School of Physics and Electronics, Hunan University, Changsha 410082, China}
 
\date{\today}
	
\begin{abstract}
The high-Chern number phases with a Chern number $C>1$ have been observed in a recent experiment that performed on the topological insulator (TI) multilayer structures, consisting of the alternating magnetic-doped and undoped TI layers.  In this paper, we develop an effective method to determine the Chern numbers in the TI multilayer structures and then make a systematic study on the Chern number phase diagrams that are modulated by the magnetic doping and the middle layer thickness.  We point out that in the multilayer structure, the high-$C$ behavior can be attributed to the band inversion mechanisms.  Moreover, we find that the lowest bands may be multifold degenerate around the $\Gamma$ point, and when they are inverted, the Chern number change will be larger than one.  Besides the TI multilayer structures implemented in the experiment, we also explore the high-$C$ phase realizations in two other kinds of the TI multilayer structures.  The implications of our results for experiments are discussed.  
\end{abstract} 

\maketitle

\section{Introduction} 

The Chern insulator, characterized by the Chern number $C$~\cite{Thouless}, was first proposed in the Haldane model~\cite{Haldane}, where the dissipationless quantized Hall transport was realized in the absence of an external magnetic fields~\cite{Qi2008}.  Its appearance is closely related to the broken time-reversal symmetry and the induced topologically nontrivial band structures~\cite{D.Xiao, M.Z.Hasan, Qi2011}.  Thanks to the development of the cold atoms, the Haldane model was simulated and demonstrated in
periodically modulated optical honeycomb lattices~\cite{G.Jotzu}.  In the real material systems, the Chern insulator has been observed successfully in two kinds of systems.  One is the magnetic-doped topological insulator (TI) thin films, as in the Cr-doped or V-doped (Bi,Sb)$_2$Te$_3$~\cite{C.Z.Chang2013, X.Kou, J.G.Checkelsky, C.Z.Chang2015a}.  Another is the currently hot topic of the twisted bilayer graphene around the magic angle, when being sandwiched between the aligned hexagonal boron nitride layers~\cite{A.L.Sharpe, M.Serlin, J.Zhu}.  In all these experiments, the Chern insulator was limited to the $C=1$ phase.  

The high-$C$ phase refers to the phase with a Chern number larger than one.  In the high-$C$ phase, although the chiral edge currents along the system's boundary are dissipationless, the contact Hall resistance between the chiral edge channels and normal metal electrodes is predicted to be $\frac{h}{Ce^2}$, according to the Landauer-B{\"u}ttiker transport theory~\cite{R.Landauer, C.Z.Chang2015b}.  As the high-$C$ phase shows potential applications in the multi-channel quantum computing~\cite{X.Zhang}, it  has aroused a lot of interests.  In the two-dimensional (2D) system, the high-$C$ phase was proposed by adding the number of the Dirac points, which could be achieved by introducing the distant hoppings beyond the next-nearest-neighboring hoppings in the Haldane model~\cite{S.Yang, D.Sticlet2012, D.Sticlet2013, Y.X.Wang2015a, Y.X.Wang2015b}.  In the twisted double bilayer graphene system, when the valley splitting is large enough to open a gap at half-filling of the first conduction band, the phases with $C=4$ and $C=2$ can be realized in the AB-AB and AB-BA stacking, respectively~\cite{Y.X.Wang2021}.  In the three-dimensional (3D) system, the high-$C$ phase was proposed in the magnetic-doped TI and magnetic-doped topological crystalline insulator thin films~\cite{J.Wang, C.Fang}.  In the former system, the high-$C$ behavior is due to the inverted bands that are induced by the strong Zeeman splitting~\cite{J.Wang, H.Jiang}, while in the latter, it is due to the presence of the multiple Dirac surface states~\cite{C.Fang}.  More interestingly, a recent study reported that the high-$C$ phases were successfully observed in the (magnetic-doped TI-undoped TI)$_m$-magnetic-doped TI multilayer structures, with the schematics plotted in Fig.~\ref{Fig1}(a)~\cite{Y.F.Zhao}.  In fact, a similar magnetic-doped TI and ordinary insulator multilayer structure was used to design the 3D Weyl semimetal phase~\cite{A.A.Burkov, A.A.Zyuzin}. 

Motivated by these experimental and theoretical progresses, in this paper, we try to address two important questions related to the TI multilayer structures: (i) what is the underlying physical mechanism of the high-$C$ phase, and (ii) besides the structure in Fig.~\ref{Fig1}(a), whether there exist other TI multilayer structures to realize the high-$C$ phase, such as those in Figs.~\ref{Fig1}(b) and (c).  In Ref.~\cite{Y.F.Zhao}, the authors answered the first question through the interface Dirac states that are spatially localized at the interfaces between the magnetic-doped TI layers and undoped ones.  They stated that the interface Dirac states were mainly determined by the undoped TI layers in the multilayer structure with each layer hosting two Dirac surface states at the $\Gamma$ point, when occupied, contributing $\frac{e^2}{2h}$ to the total Hall conductance $\sigma_{xy}$.  For the undoped TI layer number $m$, there exist $2m$ gapped interface Dirac bands that can add up to $C=m$.  This picture can explain their experimental observations at $m=2$ and the weak Zeeman splitting~\cite{Y.F.Zhao}, but may not be applicable to case of the $C>m$ phase at strong Zeeman splitting, which requires that the number of the interfaces be larger than $2m$ for a fixed multilayer structure with maximum $2m$ interfaces. 

Here we will develop an effective method to determine the Chern number in the TI multilayer structures and then make a systematic study on the Chern number phase diagrams that are modulated by the magnetic doping and the middle layer thickness.  For the first question, our calculations suggest that the high-$C$ phase could be more appropriately explained through the band inversion mechanisms, in which when the bulk gap gets closed and reopen, the wavefunction components are exchanged at the inversion point and as a result, the Chern number changes.  When multiple band inversions occur, the high-$C$ number phase will form.  Compared with the magnetic-doped TI thin film, the multifold band degeneracy around the $\Gamma$ point can be found in the TI multilayer structure and when they are inverted, the Chern number change will be larger than one.  It should be emphasized that although the band inversion mechanisms and the interface-Dirac-state mechanisms are both bulk properties, they are distinct from each other, where the former may occur for all low-energy bands, while the latter only refer to the $2m$ lowest occupied bands.  As a result, the band inversions are not directly related to the change of the number of interface Dirac bands.  For example, in both Fig. 4(c) with $C=1$ and Fig. 4(d) with $C=3$, the six lowest bands are all localized on the interfaces and thus cannot be used to explain the Chern number phase transition.  Similar cases also occur in Fig. 6(c) with $C=2$ and Fig. 6(d) with $C=3$.  For the second question, our calculations show that the other two kinds of the TI multilayer structures can also support the high-$C$ phases, but their bulk gaps are too small and thus will hinder the high-$C$ observations in the experiment.  We hope that our studies could help understand the high-$C$ behaviors in the TI multilayer structures and pave the way for their applications in the future topological electronic devices.

\section{Model and Method}

We start from the model for a three-dimensional (3D) magnetic-doped TI thin film along the $z$ axis~\cite{H.Zhang, C.X.Liu}.  The low-energy excitations around the $\Gamma$ point consist of a bonding and antibonding state of the $p_z$ orbitals, labeled by $|P1_z^+,\uparrow(\downarrow)\rangle$ and $|P2_z^-,\uparrow(\downarrow)\rangle$, with $\pm$ being the even and odd parity, and $\uparrow$, $\downarrow$ denoting the upspin and downspin.  In the four-component basis $\begin{pmatrix} 
|P1_z^+,\uparrow\rangle& 
|P2_z^-,\downarrow\rangle& 
|P1_z^+,\downarrow\rangle& 
|P2_z^-,\uparrow\rangle
\end{pmatrix}^T$, the Hamiltonian is written as~\cite{H.Zhang, C.X.Liu}
\begin{align}
&H(k_x,k_y,z)=\begin{pmatrix}
H_+(k_x,k_y,z)& -iB\partial_z\tau_y
\\
-iB\partial_z\tau_y& H_-(k_x,k_y,z)
\end{pmatrix}, 
\label{Hamil}
\end{align}
where $H_\pm(k_x,k_y,z)=\varepsilon(k_x,k_y,z)+Ak_y\tau_x\mp Ak_x\tau_y+[M(k_x,k_y,z)\pm g]\tau_z$, $\tau$ is the Pauli matrice, and $B$ couples $H_+$ and $H_-$.  To the lowest order in the wave vector, 
$\varepsilon(k_x,k_y,z)=D_0+D_1(-\partial_z^2)+D_2(k_x^2+k_y^2)$ accounts for the particle-hole asymmetry and $M(k_x,k_y,z)=M_0'+M_1 (-\partial_z^2)+M_2(k_x^2+k_y^2)$.  The effects of magnetic Cr-doping in a TI thin film are twofolds~\cite{Y.F.Zhao}: (i) It reduces the spin-orbit coupling of the system and drives their bulk energy gap towards the non-inverted insulator regime, which is captured by the bulk gap of the magnetic-doped TI $M_0'$.  (ii) It introduces the magnetic moments in the system, resulting in a Zeeman splitting, which is represented by $g$. 

\begin{figure}
	\includegraphics[width=8.4cm]{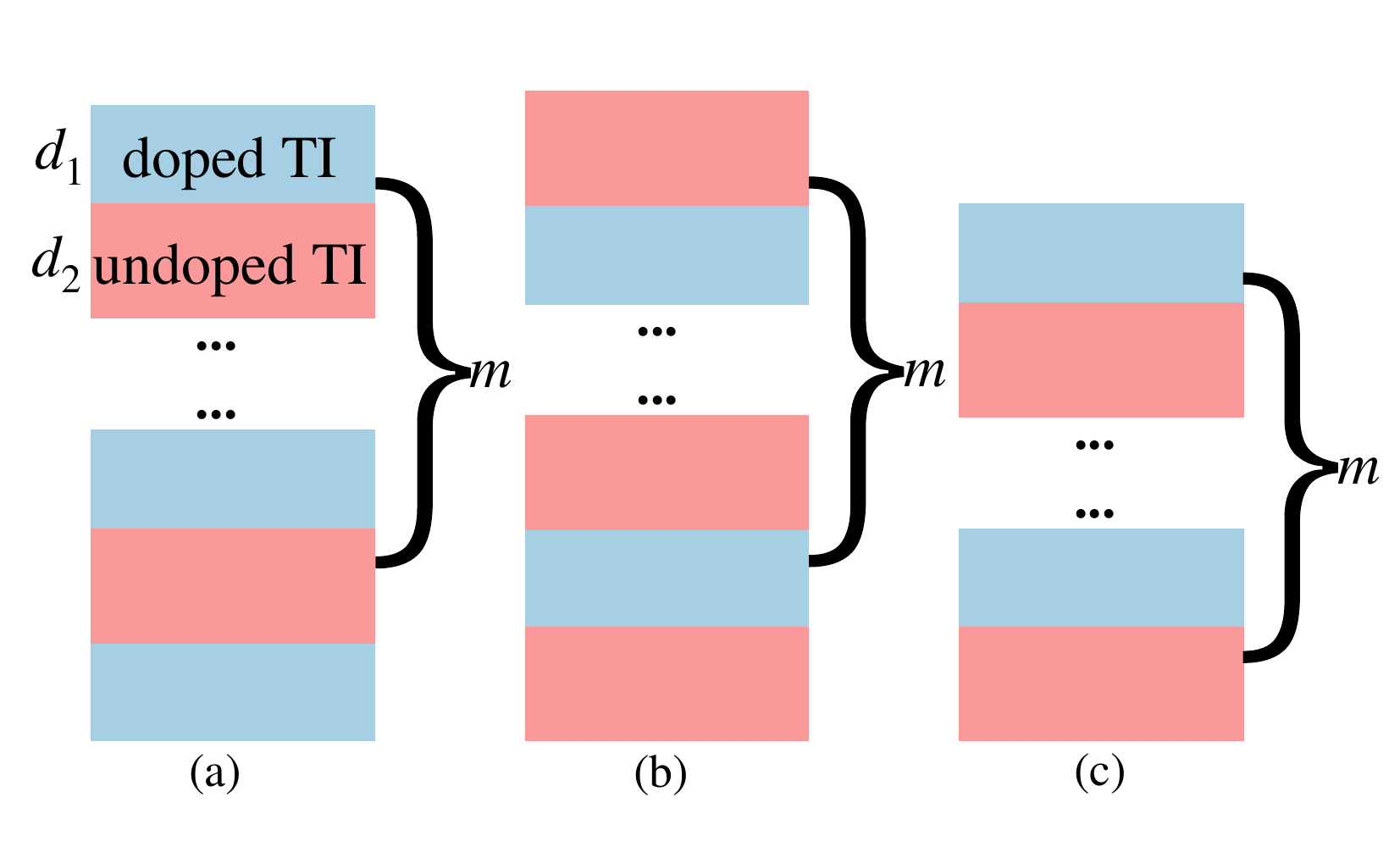}
	\caption{Schematics of the TI multilayer structures, which consist of the alternating magnetic-doped TI layer and undoped TI layer, with (a) the (doped TI-undoped TI)$_m$-doped TI multilayer, (b) the (undoped TI-doped TI)$_m$-undoped TI multilayer, and (c) the (doped TI-undoped TI)$_m$ multilayer.  We use $m$ to denote the  double-layer unit number and set $d_1=3$ nm and $d_2=4$ nm as the thickness of the doped TI layer and undoped TI layer, respectively.  We choose the $z$ direction as the stacking direction and the structure center as the origin. }
	\label{Fig1}
\end{figure}

In Fig.~\ref{Fig1}, the schematics of the TI multilayer structures are plotted, consisting of the alternating magnetic-doped TI layer and undoped TI layer.  We use $m$ to denote the double-layer unit number and set $d_1=3$ nm and $d_2=4$ nm as the thickness of the doped TI layer and undoped TI layer, respectively.  The multilayer structures can also be described by the Hamiltonian in Eq.~(\ref{Hamil}), but with $M_0'$ replaced by 
\begin{align}
{\cal M}_0(z)=\left\{\begin{array}{l}
M_0  \quad (z\in\text{undoped TI layer})
\\
M_0'  \quad (z\in\text{Cr-doped TI layer})
\end{array} \right., 
\end{align}
and $g$ replaced by 
\begin{align}
g(z)=\left\{ \begin{array}{l}
0  \quad (z\in\text{undoped TI layer})
\\
g  \quad (z\in\text{Cr-doped TI layer})
\end{array}\right..
\end{align}

The confinement of a finite thickness $L$ in the multilayer structure can quantize the momentum in the $z$ axis, leading to 2D subbands.  If the multilayer structure center is set as the origin, we can choose $\varphi_n(z)=\sqrt{\frac{2}{L}}
\text{sin}(\frac{n\pi z}{L}+\frac{n\pi}{2})$ to satisfy the open boundary conditions at $z=\pm\frac{L}{2}$.  The open boundary condition is also used in the study of the effect of finite size on the edge states in a quantum spin Hall system~\cite{B.Zhou}.  We then express the Hamiltonian in the $n-$subband basis $\begin{pmatrix}
\varphi_n|P1_z^+,\uparrow\rangle& \varphi_n|P2_z^-,\downarrow\rangle& 
\varphi_n|P1_z^+,\downarrow\rangle&  \varphi_n|P2_z^-,\uparrow\rangle
\end{pmatrix}^T$.  

Consider the TI multilayer structure in Fig.~\ref{Fig1}(a).  When the Zeeman splitting is absent, $g=0$, the Hamiltonian possesses the time-reversal symmetry as ${\cal T}H(-k_x,-k_y,z){\cal T}^{-1}=H(k_x,k_y,z)$, where the time-reversal operator is
${\cal T}=\begin{pmatrix}
0& \tau_zK\\
-\tau_zK& 0
\end{pmatrix}$, with $K$ being the complex conjugate operator.  The introduction of the Zeeman splitting will break the time-reversal symmetry.  The Hamiltonian also owns the twofold rotational symmetry along the $z$ axis, ${\cal C}_{2z}^{-1} H(-k_x,-k_y,z){\cal C}_{2z}=H(k_x,k_y,z)$, the mirror symmetry with the $x-y$ plane, ${\cal M}^{-1} H(k_x,k_y,-z){\cal M}=H(k_x,k_y,z)$, and the inversion symmetry, 
${\cal P}^{-1} H(-k_x,-k_y,-z){\cal P}=H(k_x,k_y,z)$, with the operators being  
${\cal C}_{2z}=\begin{pmatrix}
\tau_z& 0\\ 0& -\tau_z\end{pmatrix}$,  
${\cal M}=\begin{pmatrix}
I& 0\\ 0& -I
\end{pmatrix}$, and  
${\cal P}={\cal C}_{2z}{\cal M}=\begin{pmatrix}
\tau_z& 0\\ 0& \tau_z
\end{pmatrix}$.  For the terms of $M_1\langle-\partial_z^2\rangle_n+M_2(k_x^2+k_y^2)$, ${\cal M}_0(z)$, $g(z)$, $B\partial_z$, and $A(k_y+ik_x)$, the corresponding matrix elements $R_{1,\cdots,5}$ can be expressed in the subband space as   
\begin{align}
R_1^{n_1,n_2}
=&\left\{ \begin{array}{l}
M_1\frac{n^2\pi^2}{L^2}+M_2(k_x^2+k_y^2)  \quad (n_1=n_2)
\\
0  \quad (n_1\neq n_2) 
\end{array} \right., 
\\
R_2^{n_1,n_2}=&M_0'f_1^{n_1,n_2}+M_0f_2^{n_1,n_2}, 
\label{R2}
\\
R_3^{n_1,n_2}=&g f_1^{n_1,n_2}, 
\label{R3}
\\
R_4^{n_1,n_2}
=&\left\{ \begin{array}{l}
0 \quad (n_1=n_2)  
\\
\frac{2Bn_1n_2}{(n_1^2-n_2^2)L} [1-(-1)^{n_1+n_2}]  \quad (n_1\neq n_2)
\end{array} \right., 
\end{align} 
and
\begin{align}
R_5^{n_1,n_2}
=&\left\{ \begin{array}{l}
A(k_y+ik_x)  \quad (n_1=n_2)
\\
0  \quad (n_1\neq n_2) 
\end{array} \right..  
\end{align}
The explicit forms of the functions $f_1^{n_1,n_2}$ and $f_2^{n_1,n_2}$ in Eqs.~(\ref{R2}) and (\ref{R3}) are given in Appendix A.  We can see that $R_1$ and $R_5$ do not depend on the TI multilayer structures and are diagonal in the subband space.  $R_2$ and $R_3$ include the diagonal terms as well as the off-diagonal ones, where the former can modulate the Dirac mass $m_{n\pm}$ and the latter depend on the TI multilayer structures.  In addition, $R_4$ only includes the off-diagonal terms.  Although the dimension of the Hamiltonian is infinity, we take the subband cutoff as $N_{\text{c}}=50$ to obtain well convergent results.  This is justified because the $n-$th component in the subband space decays rapidly when $n$ is beyond a certain value.  

To determine the Chern number in the multilayer structures, we can adopt the following steps: 

(i) Judge the Chern number in the special case when all the off-diagonal terms of $R_2$, $R_3$, and $R_4$ are absent and only the diagonal terms of $R_1$, $R_2$, $R_3$, and $R_5$ are kept; 

(ii) Adiabatically add the off-diagonal terms of $R_2$ and $R_3$ to the system;  

(iii) Slowly increase the coupling $B$ from zero to its realistic value.  

In step (i), when all off-diagonal terms are neglected, the diagonal Hamiltonian is decoupled into two classes of 2D models, $h_+(n)$ and $h_-(n)$ with opposite chiralities, 
\begin{align}
&\tilde H_\text{2D}(k_x,k_y,n)=\begin{pmatrix}
h_+(k_x,k_y,n)& 0
\\
0& h_-(k_x,k_y,n)
\end{pmatrix},
\\
&h_{\pm}(k_x,k_y,n)=\tilde\varepsilon_n+Ak_y\tau_x\mp Ak_x\tau_y
+m_{n\pm}\tau_z, 
\end{align}
where $\tilde\varepsilon_n=D_0+D_1\langle-\partial_z^2\rangle_n+D_2(k_x^2+k_y^2)$ and $m_{n\pm}=R_1^n+R_2^n\pm R_3^n$ is the Dirac mass for the $n\pm$ subband.  When setting $\tilde\varepsilon_n=0$ and pinning the system at half-filling, the particle-hole symmetry is preserved.  In this case, the 2D models $h_\pm(n)$ have a Chern number $\pm1$ or $0$, depending on whether the Dirac mass at the $\Gamma$ point is inverted, $m_{n\pm}<0$, or non-inverted, $m_{n\pm}>0$.  As a result, the total Chern number of the system is~\cite{J.Wang}
\begin{align}
C=N_+-N_-,
\label{Chern}
\end{align}
where $N_\pm$ is the number of $h_\pm(n)$ with the inverted Dirac mass, respectively.  In steps (ii) and (iii), the evolutions of the phase boundaries need to be captured, as the Chern number change is related to the bulk gap closings of the system. 

In the following, we use the parameters in Bi$_2$Se$_3$~\cite{C.X.Liu}: $M_0=-0.28$ eV, $M_1=0.0686$ eV$\cdot$nm$^2$, $M_2=0.445$ eV$\cdot$nm$^2$, $A=0.333$ eV$\cdot$nm and $B=0.226$ eV$\cdot$nm.

\section{magnetic-doped TI thin film}

\begin{figure}
	\includegraphics[width=9cm]{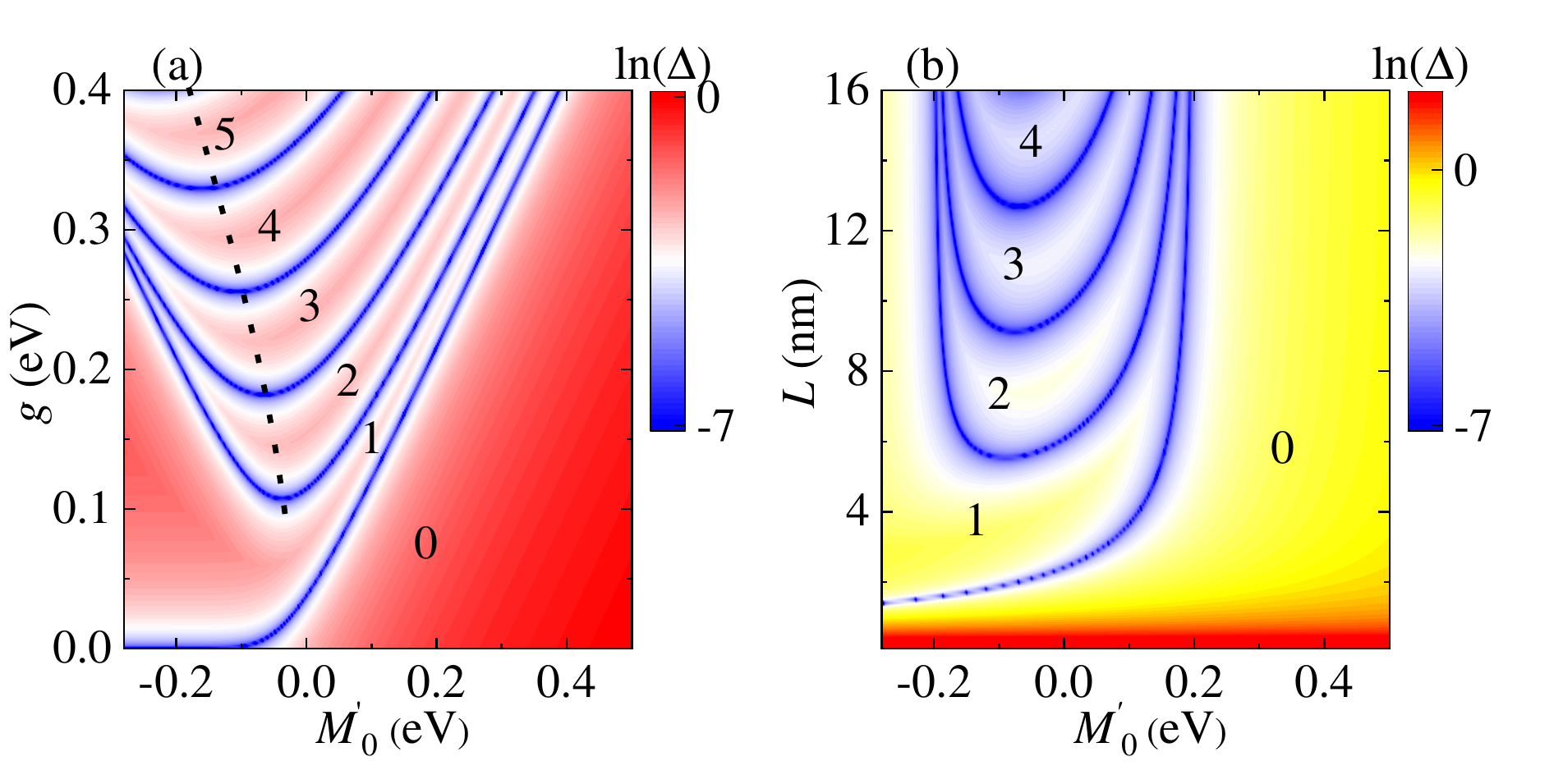}
	\caption{Chern number phase diagrams of the magnetic-doped TI thin film, with the Chern number values being labeled.  (a) is plotted in the parameter space of $M_0'$ and Zeeman splitting $g$, and (b) is plotted in the parameter space of $M_0'$ and the thin film thickness $L$.  The contour scale represents the magnitude of ln$(\Delta)$, where $\Delta$ is the energy gap of the lowest bands.  The bright blue lines denote that the gap is closed and separate the different Chern number phase.  The dotted line in (a) separates the regions where the Chern number increases or decreases with $M_0'$.  We choose $L=10$ nm in (a) and $g=0.2$ eV in (b). }
	\label{Fig2}
\end{figure}

\begin{figure*}
	\centering
	\includegraphics[width=18.4cm]{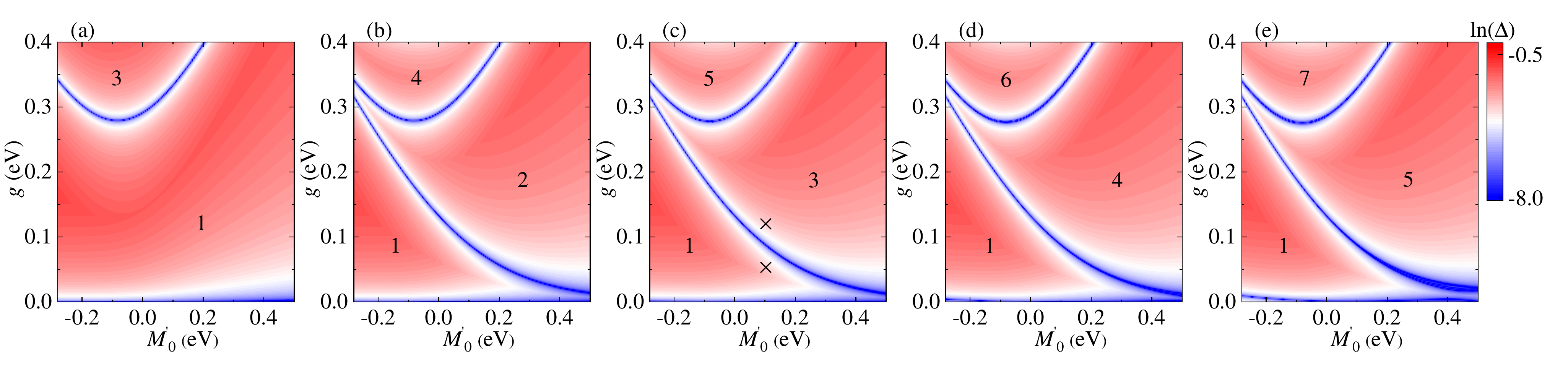}
	\caption{Chern number phase diagrams of the (doped TI-undoped TI)$_m$-doped TI multilayer structures in the parameter space of $M_0'$ and Zeeman splitting $g$, with the Chern number values being labeled.  From left to right, $m$ increases from $1$ to $5$.  The contour scale represents the magnitude of ln$(\Delta)$, where $\Delta$ is the energy gap of the lowest bands.  The bright blue lines indicate that the gap is closed and separate the different Chern number phases. }
	\label{Fig3}
\end{figure*}

Intuitively, the appearance of the nontrivial Chern number phase is determined by the competition between the magnetic-doped and undoped TI layers.  We first present the Chern number phase diagrams of a magnetic-doped TI thin film in Fig.~\ref{Fig2}.  The contour scale represents the magnitude of ln$(\Delta)$, with $\Delta$ being the energy gap of the lowest bands.  We can also use the above method to determine the Chern number in a TI thin film, but with step (ii) being absent, as $R_2$ and $R_3$ are diagonal in the subband space here.  Compared with Ref.~\cite{J.Wang} of the weak coupling $B$ value, where the regions of the odd Chern number phase are connected and those of the even Chern number phase are separated into islands, here in Fig.~\ref{Fig2}, we find that the regions of the same Chern number phase are all connected due to the strong coupling $B$ value.  

Figure~\ref{Fig2}(a) is plotted in the parameter space of $M_0'$ and Zeeman splitting $g$.  When the Zeeman splitting $g=0$, the time-reversal symmetry is preserved and the Dirac masses for $n\pm$ subbands are equal.  So $n\pm$ subbands are inverted or non-inverted simultaneously, leading to the equivalence of $N_+$ and $N_-$ in Eq.~(\ref{Chern}) and the Chern number $C=0$.  
However, a tiny magnetic-doping will drive the TI thin film from the trivial $C=0$ phase to enter the nontrivial $C=1$ phase.  We can also see that the Chern number increases with $g$, but exhibits a non-monotonous behavior with $M_0'$.  When the dotted line is crossed, the Chern number decreases with $M_0'$.  This is because the growing of $g$ can change the Dirac mass and thus drive more subband gaps into the inverted regime, while the increasing of $M_0'$ can reduce the spin-orbit coupling of the system, which thus cause the subband gap to become non-inverted.  Figure~\ref{Fig2}(b) is plotted in the parameter space of $M_0'$ and the thin film thickness $L$, where the Chern number increases with $L$, as more subband gaps will get inverted in a thicker sample.  In both Figs.~\ref{Fig2}(a) and (b), each time a phase boundary is crossed, the Chern number will be changed by one. 

According to the above results, the necessary conditions for the appearance of the high-$C$ phase in a magnetic-doped TI thin film include the strong $g$, low $M_0'$.  In addition, the thickness $L$ cannot be too small, \textit{e.g.}, in Fig.~\ref{Fig2}(b) when $g=0.2$ eV, $L$ cannot not be smaller than 5.5 nm.  These conditions are consistent with the previous studies~\cite{J.Wang, H.Jiang}.  However, in the experiment, the magnetic doping in a TI thin film can make $g$ and $M_0'$ grow simultaneously, which thus break the necessary conditions.  This explains why only $C=1$  phase was observed in a Cr-doped (Bi,Sb)$_2$Te$_3$ TI thin film sample with $L=5$ nm~\cite{C.Z.Chang2013}, 8 nm~\cite{J.G.Checkelsky}, and 10 nm~\cite{X.Kou}.  Therefore we arrive at the conclusion that the high-$C$ phase is difficult to be observed in a magnetic-doped TI thin film.  But it can appear in the TI multilayer structures, as we explain below.

\section{(Doped TI-undoped TI)$_m$-doped TI multilayer structures} 

\begin{figure}
	\includegraphics[width=9cm]{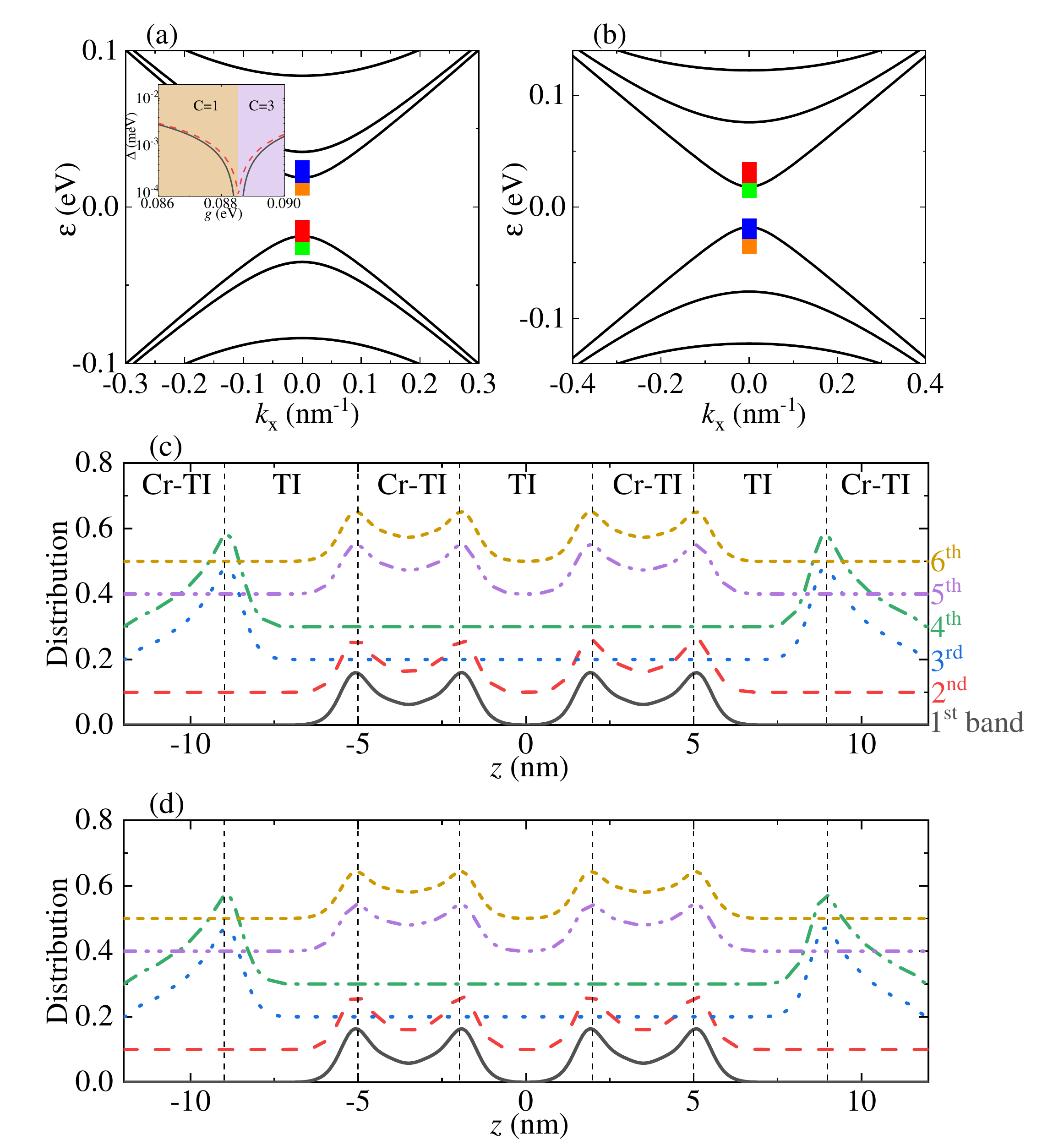}
	\caption{The six lowest energy bands and the normalized wavefunction distributions at the $\Gamma$ point in the $m=3$ multilayer structure.  We choose the parameters as $g=0.056$ eV in (a) and (c), $g=0.12$ eV in (b) and (d), and $M_0'=0.1$ eV, as labeled by the crosses in Fig.~\ref{Fig3}(c).
	In (a) and (b), all bands are twofold degenerate around the $\Gamma$ point.   
	The orbital contributions at the $\Gamma$ point to the inverted bands are shown by the colored rectangles, where the red, green, blue and orange colors denote the $|P1_z^+,\uparrow\rangle$, $|P2_z^-,\uparrow\rangle$, $|P2_z^-,\downarrow\rangle$, and $|P1_z^+,\downarrow\rangle$ orbitals, respectively, and the heights are proportional to the weights of the orbitals.  The inset in (a) gives evolution of the gaps for the two lowest bands with $g$ around the critical point $g_c=0.0885$ eV. 
	In (c) and (d), the vertical dotted lines denote the interfaces between the doped TI and the undoped TI and the legends are the same.  For clarity, the neighboring lines  are vertically shifted by 0.1. }
	\label{Fig4}
\end{figure}

\begin{figure*}
	\centering
	\includegraphics[width=18.4cm]{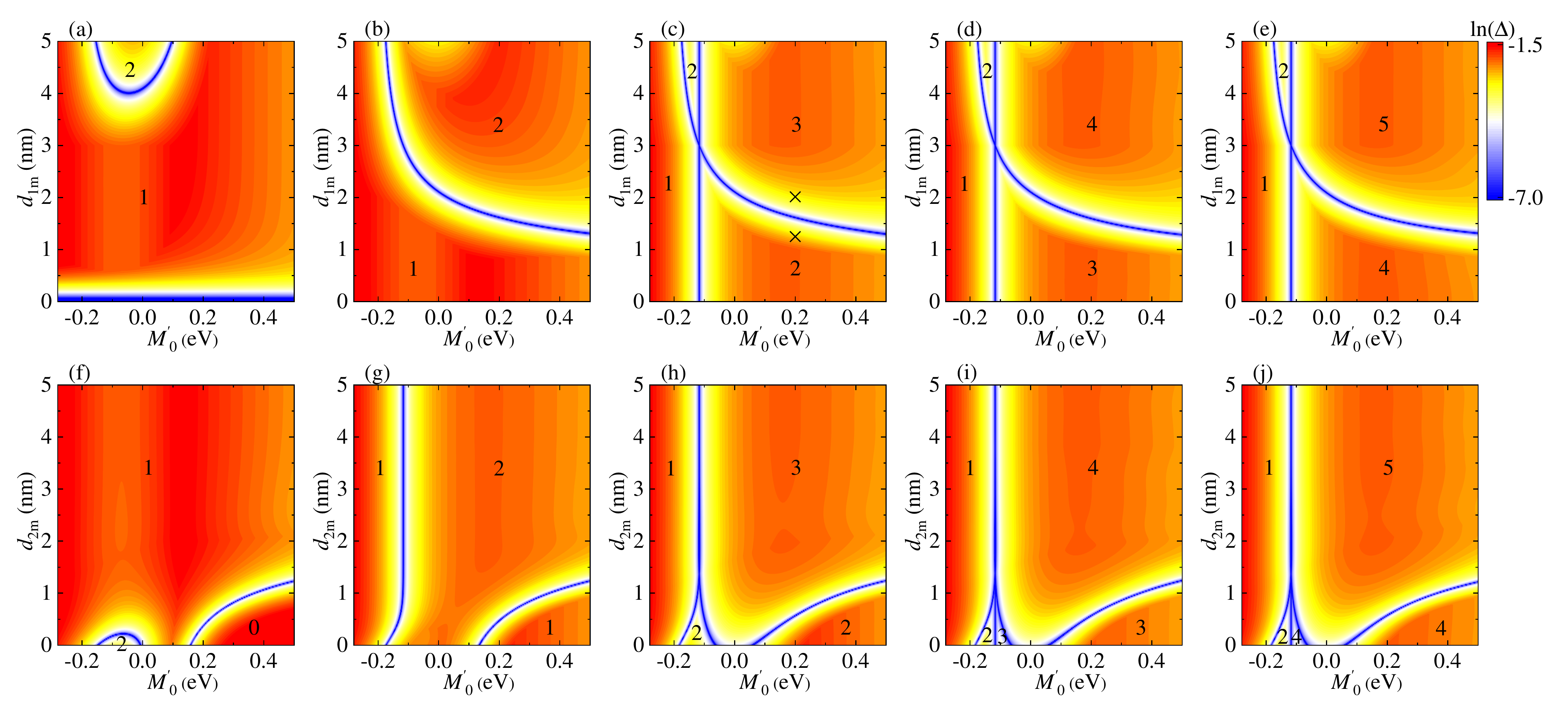}
	\caption{Chern number phase diagrams of the (doped TI-undoped TI)$_m$-doped TI multilayer structures in the parameter space of $M_0'$ and the middle layer thickness, with the Chern number values being labeled.  In (a)$-$(e), $d_{1m}$ denotes the thickness of the middle doped TI layer or the doped TI layer nearest to the middle layer, while in (f)$-$(j), $d_{2m}$ denotes the thickness of the middle undoped TI layer or the undoped TI layer nearest to the middle layer.  From left to right, $m$ increases from $1$ to $5$.  The contour scale represents the magnitude of ln$(\Delta)$, where $\Delta$ is the energy gap of the lowest bands.  The bright blue lines indicate that the gap is closed and separate the different Chern number phases.  We choose the Zeeman splitting $g=0.2$ eV. }
	\label{Fig5}
\end{figure*}

Next we study the TI multilayer structures that were implemented in the experiment~\cite{Y.F.Zhao}.  The steps (i)$-$(iii) are used to determine the Chern number phase diagram in the system.  In step (i) when all the off-diagonal terms are absent, the phase boundaries are given by the zeros of the Dirac mass of $n\pm$ subbands, $m_{n\pm}=0$, which correspond to
\begin{align}
&M_0'-M_0+\big(M_0+M_1\frac{n^2\pi^2}{L^2}\big)
\Big[\frac{(m+1)d_1}{L}+\frac{1}{n\pi}\text{sin}\big(\frac{d_2 n\pi}{L}\big)
\nonumber\\
&
\times\sum_{i=1}^m\text{cos}\Big(\frac{(2i d_1+2i d_2-d_2)n\pi}{L}
\Big)\Big]^{-1}\pm g=0. 
\label{phaseboundary}
\end{align}
When the Zeeman splitting $g=0$, the time-reversal symmetry is preserved and thus the Chern number also vanishes, which is the same as the TI thin film.  In the experiment~\cite{Y.F.Zhao}, it was demonstrated that the Chern number phases can be effectively tuned from $C=1$ to $C=2$ in the $m=2$ TI multilayer structures by varying either the Cr-doping concentration in (Bi,Sb)$_{2-x}$Cr$_x$Te$_3$ from $x=0.13$ to $x=0.35$ or the middle doped TI layer thickness from $d=0$ to $d=4$ nm.  So we will study the Chern number phase diagram modulated by the magnetic doping as well as the middle layer thickness in the system. 

\begin{figure}
	\includegraphics[width=9cm]{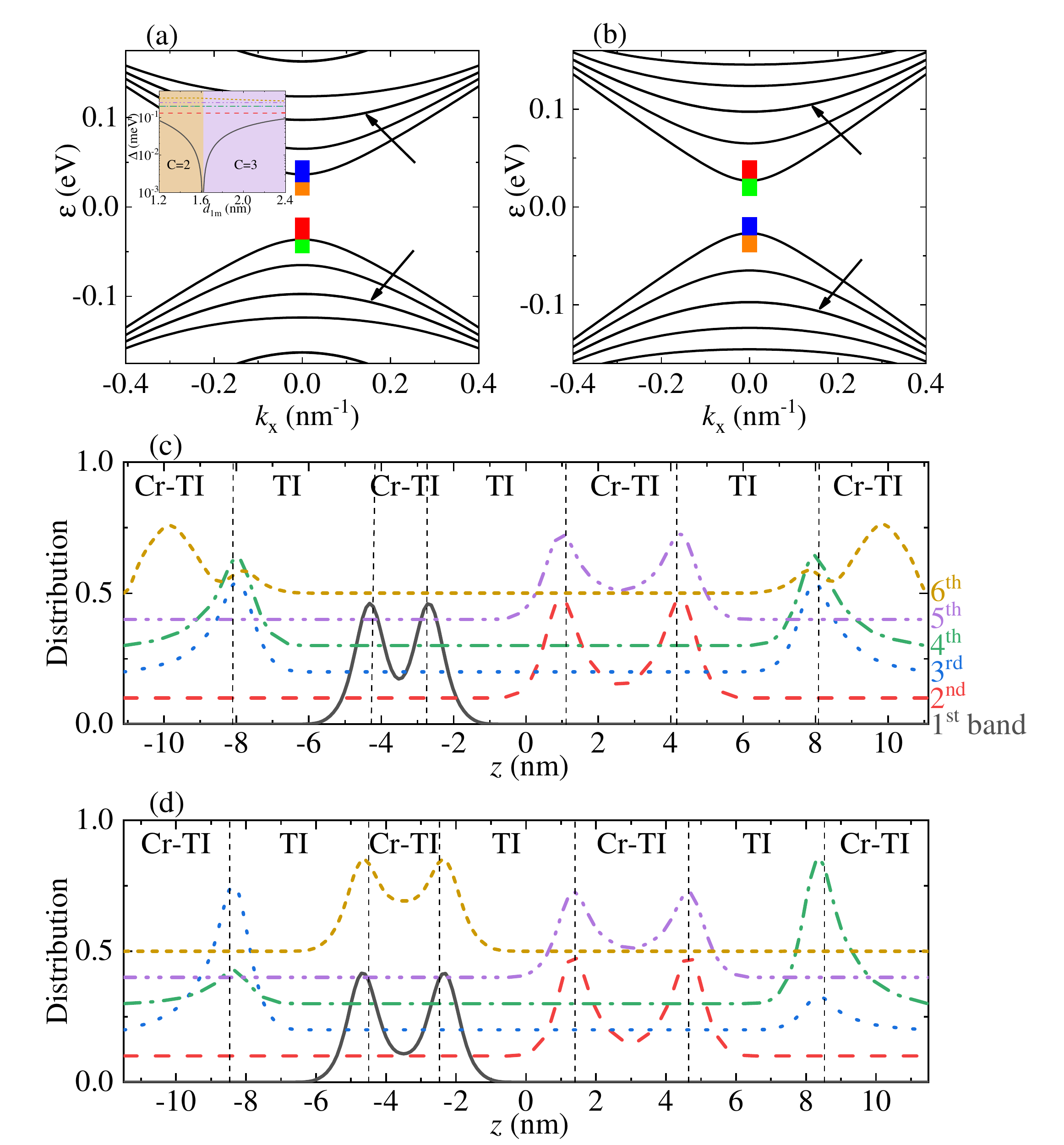}
	\caption{The six lowest energy bands and the normalized wavefunction distributions at the $\Gamma$ point in the $m=3$ TI multilayer structure.  We choose the parameters as $d_{1m}=1.24$ nm in (a) and (c), $d_{1m}=2$ nm in (b) and (d), and $M_0'=g=0.2$ eV, as labeled by the crosses in Fig.~\ref{Fig5}(c). 
	In (a) and (b), the third and fourth conduction/valence bands are degenerate around the $\Gamma$ point, as labeled by the arrows, while other bands are non-degenerate.  
	The orbital contributions at the $\Gamma$ point to the inverted bands are shown by the colored rectangles, where the red, green, blue and orange colors denote the $|P1_z^+,\uparrow\rangle$, $|P2_z^-,\uparrow\rangle$, $|P2_z^-,\downarrow\rangle$, and $|P1_z^+,\downarrow\rangle$ orbitals, respectively, and the heights are proportional to the weights of the orbitals.  The inset in (a) gives evolution of the gaps for the six lowest bands with $d_{1m}$ around the critical point $d_{1mc}=1.62$ nm. 
	In (c) and (d), the vertical dotted lines denote the interface between the doped TI and the undoped TI and the legends are the same.  For clarity, the neighboring lines  are vertically shifted by 0.1. }
	\label{Fig6}
\end{figure}

In Fig.~\ref{Fig3}, the Chern number phase diagrams of the multilayer structures are plotted in the parameter space of $M_0'$ and Zeeman splitting $g$, with the double-layer unit number increasing from $m=1$ to $m=5$.  An example of Fig.~\ref{Fig3}(c) can be found in Appendix B with more details about the Chern number determination.  We can see that the phase diagrams exhibit similar structures and include the phases of $C=0,1,m,m+2$.  Compared to the TI thin film, here the high-$C$ phase can appear in the region of large $M_0'$ and low $g$.  In Figs.~\ref{Fig3}(a)$-$(e), the region spanned by the $C=0$ phase is vanishing, as the phase boundary that separates the $C=0$ and $C=1$ phases is close to the $g=0$ line, meaning that a tiny Zeeman splitting caused by the magnetic doping can drive the system from the trivial $C=0$ phase to the nontrivial $C=1$ phase.  In the middle region, no phase boundary exists in Fig.~\ref{Fig3}(a), where only the $C=1$ phase is present, while in Figs.~\ref{Fig3}(b)$-$\ref{Fig3}(e), the phase boundary separates the $C=1$ and $C=m$ phases.  In addition, the upper phase boundary separates the $C=m$ and $C=m+2$ phases.  Note that Fig.~\ref{Fig3}(b) is consistent with Fig.~4(d) in Ref.~\cite{Y.F.Zhao} obtained by using the Kubo formula, except that it is the $C=4$ phase that appears at strong $g$, but not $C=3$ as labeled in Ref.~\cite{Y.F.Zhao}. 

In Fig.~\ref{Fig4}, we plot the six lowest energy bands and the normalized wavefunction distributions at the $\Gamma$ point in the $m=3$ TI multilayer structure, with the chosen parameters being labeled as the crosses in Fig.~\ref{Fig3}(c).  By exactly counting the band data, we find that in Figs.~\ref{Fig4}(a) and (b), all bands are twofold degenerate around the $\Gamma$ point, which can be attributed to the specific TI multilayer structure.  In Figs.~\ref{Fig4}(c) and (d), the wavefunctions are symmetric to $z=0$ and are nearly overlapped for the degenerated bands.  More importantly, the wavefunctions show minor variations when the system changes from the $C=1$ phase to $C=3$, where the wavefunctions for the third and fourth bands are localized on two outmost interfaces, while those for the other four bands are localized on the inner interfaces.  These results strongly indicate that the high-$C$ phase may not be well accounted for by the interface Dirac bands~\cite{Y.F.Zhao}. 

Instead, we try to explain the high-$C$ behavior from the band inversions.  The degenerated first and second conduction/valence bands around the $\Gamma$ point can be well distinguished in the enlarged plots of the evolution of the band gaps around the critical point $g_c=0.0885$ eV, as presented in the inset of Fig.~\ref{Fig4}(a).  When $g$ grows to cross $g_c$, the gaps for the two lowest bands will get closed and reopen, resulting in the Chern number change from $C=1$ to $C=3$. 
In Figs.~\ref{Fig4}(a) and (b), the orbital contributions to the inverted bands at the $\Gamma$ point are represented by the rectangles, where the different colors are used to label the different orbitals and the heights are proportional to the weights of the orbitals.  For the degenerated bands, the orbital contributions are almost equivalent.  In Fig.~\ref{Fig4}(a) of the $C=1$ phase, the contributions to the first conduction band are the mixed $0.64~|P1_z^+,\uparrow\rangle+0.36~|P2_z^-,\uparrow\rangle$ and those to the first  valance band are $0.64~|P2_z^-,\downarrow\rangle+0.36~|P1_z^+,\downarrow\rangle$, while in Fig.~\ref{Fig4}(b) of the $C=3$ phase, the contributions to the first conduction band are inverted as the mixed $0.575~|P2_z^-,\downarrow\rangle+0.425~|P1_z^+,\downarrow\rangle$ and those to the first valance band are inverted as $0.575~|P1_z^+,\uparrow\rangle+0.425~|P2_z^-,\uparrow\rangle$.  So the band inversions and their degeneracy can play the decisive roles in driving the high-$C$ phase in the TI multilayer structure.  In Fig.~\ref{Fig3}, the inverted bands crossing the middle phase boundary are $(m-1)$-fold degenerate, while those crossing the upper boundary are twofold degenerate. 

In Fig.~\ref{Fig5}, the Chern number phase diagrams are plotted in the parameter space of $M_0'$ and the middle layer thickness, with $m$ also increasing from 1 to 5 and the Zeeman splitting $g=0.2$ eV.  Note that when $m$ is odd (even), the middle layer in the TI multilayer structure is the undoped (doped) TI.  In  Figs.~\ref{Fig5}(a)$-$\ref{Fig5}(e), we use $d_{1m}$ to denote the thickness of the middle doped TI layer or the doped TI layer nearest to the middle layer, and in Figs.~\ref{Fig5}(f)$-$\ref{Fig5}(j), we use $d_{2m}$ to denote the thickness of the middle undoped TI layer or the undoped TI layer nearest to the middle layer.  We can see that the phase diagrams also evolve with $m$ and exhibit similar structures: (i) They all include the phases of $C=1,2,m-1,m$, suggesting that the modulation of the middle doped or undoped TI layer thickness has a certain equivalence in driving the high-$C$ phase.  (ii) A vertical phase boundary occurs at $M_0'=-0.116$ eV.  In Figs.~\ref{Fig5}(c)$-$(e), the vertical phase boundary separates the $C=2$ and $C=m$ phases when $d_{1m}>3$ nm and separates the $C=1$ and $C=m-1$ phases when $d_{1m}<3$ nm, so the inverted bands are $(m-2)$-fold degenerate.
For comparison, in Figs.~\ref{Fig5}(h)$-$(j), when $d_{2m}>1$ nm, the vertical phase boundary separates the $C=1$ and $C=m$ phases and thus the inverted bands are $(m-1)$-fold degenerate, and when $d_{2m}<1$ nm, it separates the $C=2$ and $C=m-1$ phases and thus the inverted bands are $(m-3)$-fold degenerate.  (iii) However, when varying $d_{1m}$ or $d_{2m}$, the Chern number is always changed by one and the inverted band is non-degenerated.  We also note that Fig.~\ref{Fig5}(b) is consistent with Fig.~4(e) in Ref.~\cite{Y.F.Zhao}. 

We further consider the energy bands and the normalized wavefunctions when modulating the middle layer thickness.  The results are shown in Fig.~\ref{Fig6}, with the chosen parameters being labeled as the crosses in Fig.~\ref{Fig5}(c).  In both Figs.~\ref{Fig6}(a) and (b), the third and fourth conduction/valence bands are degenerate around the $\Gamma$ point and other bands are non-degenerate.  The evolutions of the band gaps around the critical $d_{1mc}=1.62$ nm are presented in the inset of Fig.~\ref{Fig6}(a).  When $d_{1m}$ grows to cross $d_{1mc}$, the gap of the lowest band will get closed and reopen, leading to the Chern number change from $C=2$ to $C=3$.  Correspondingly, the orbital contributions to the lowest bands at the $\Gamma$ point are also inverted, as shown in Figs.~\ref{Fig6}(a) and (b).  For the wavefunctions in Figs.~\ref{Fig6}(c) and (d), they are not symmetric to $z=0$, due to the modulation of $d_{1m}$ in the TI multilayer structure.  More importantly, in Fig.~\ref{Fig6}(c) of the $C=2$ phase, the wavefunction for the sixth band has penetrated into the interiors of the top/bottom doped TI layers and the other ones are localized on the interface, while in Fig.~\ref{Fig6}(d) of the $C=3$ phase, all the wavefunctions are localized on the interfaces.  The observation of five Dirac bands localized on the interfaces in the $C=2$ phase also contradicts with the previous explanation~\cite{Y.F.Zhao}, as it assumes that the $C=2$ phase only require four Dirac bands to be localized on the interfaces.

\section{Other multilayer structures}

\begin{figure}
	\includegraphics[width=9cm]{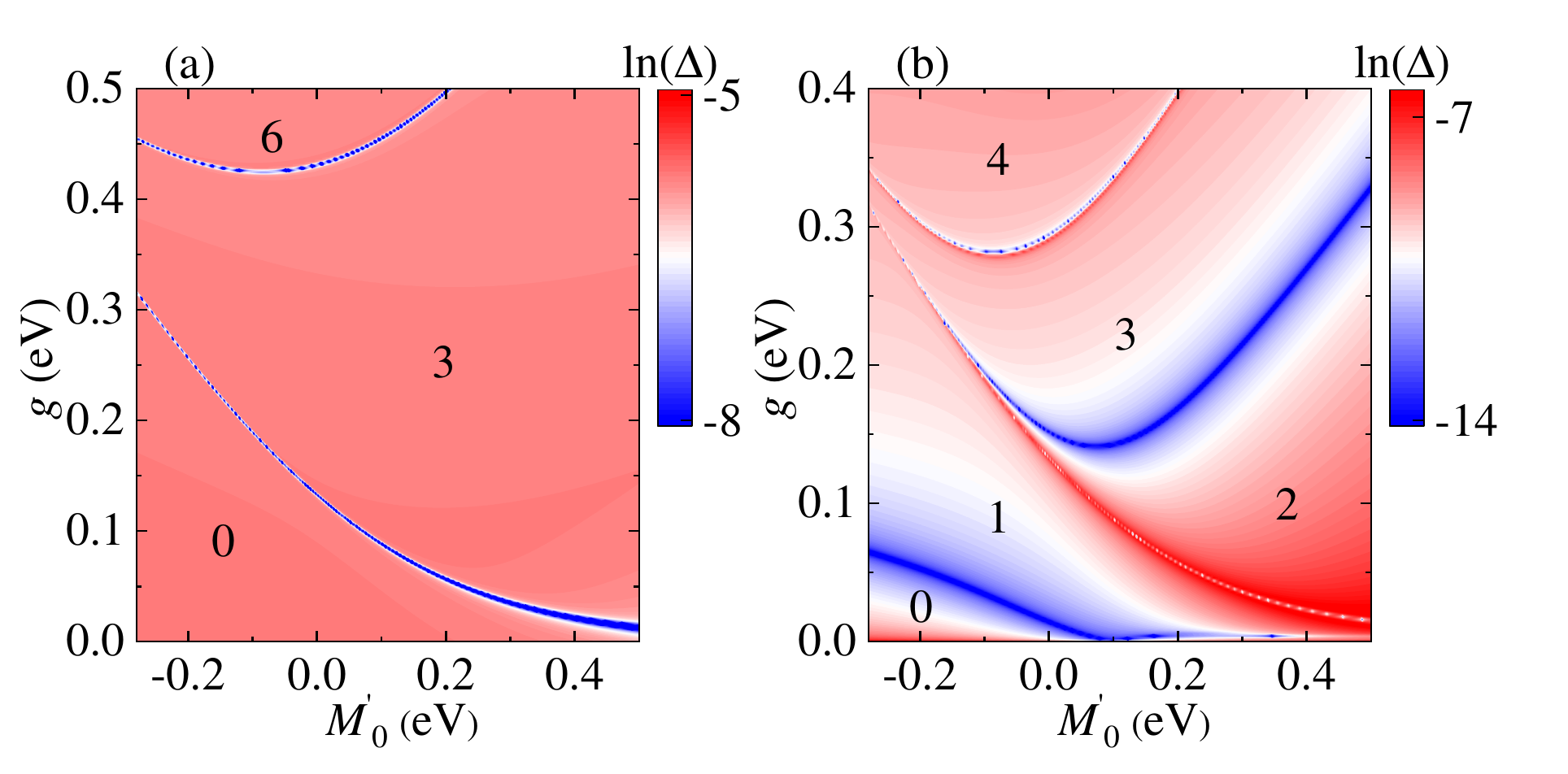}
	\caption{Chern number phase diagrams of (a) the (undoped TI-doped TI)$_{m=3}$-undoped TI and (b) the (doped TI-undoped TI)$_{m=3}$ multilayer structures in the parameter space of $M_0'$ and Zeeman splitting $g$, with the Chern number values being labeled.  The contour scale represents the magnitude of ln$(\Delta)$, where $\Delta$ is the energy gap of the lowest bands.  The bright blue lines denote that the gap is closed and separate the different Chern number phase. }
	\label{Fig7} 
\end{figure}

In this section, we explore whether there exist other TI multilayer structures to accommodate the high-$C$ phases.  We consider the structures in Figs.~\ref{Fig1}(b) and (c) and show their Chern number phase diagrams with $m=3$ in Fig.~\ref{Fig7}. 
For the structures in Fig.~\ref{Fig1}(b), the Hamiltonian satisfies all the same symmetries as Fig.~\ref{Fig1}(a), while for Fig.~\ref{Fig1}(c), the time-reversal symmetry under no Zeeman splitting, the mirror symmetry and inversion symmetry are all broken.  However, in Fig.~\ref{Fig7}(b), when $g=0$ and $M_0'=-0.28$ eV, the time-reversal symmetry is still preserved in the system and thus $C=0$, as the doped TI layer and undoped TI layer are equivalent.  So the Chern number phase diagram can also be obtained by capturing the phase boundary evolutions. 

Clearly, the high-$C$ phases can also be found in Fig.~\ref{Fig7}.  In Fig.~\ref{Fig7}(a), we can see that the $C=3$ phase spans a similar region as that in Fig.~\ref{Fig3}(c), while the higher $C=6$ phase appears at the strong Zeeman splitting with $g>0.4$ eV.  So the inverted bands crossing these two phase boundaries are both three-fold degenerate.  In Fig.~\ref{Fig7}(b), it shows that besides the Chern number phases of $C=0$ and 3, the additional $C=1$, 2 and 4 phases also appear.  

In both Figs.~\ref{Fig7}(a) and (b), the $C=0$ phase spans the lower-left region of the phase diagram and is quite different from Fig.~\ref{Fig3}, where the $C=0$ region is vanishing.  This means that the low magnetic-doping of the TI layer in these two structures will not drive the Chern number phase transitions.  Moreover, the maximum bulk gaps are only about $\Delta_m\sim e^{-5}$ in Fig.~\ref{Fig7}(a) and $\Delta_m\sim e^{-7}$ in Fig.~\ref{Fig7}(b).  
Note that in Fig.~\ref{Fig7}(b), for the transition from $C=0$ to $C=1$, as well as from $C=2$ to $C=3$, the band inversions are much slower.  These results suggest that at low-doping, when the undoped layer number is equal to or more than the doped layer number, the low-energy physics is mainly determined by the undoped layer.  In this case, the $C=0$ phase is favored and the bulk gap is almost closing.  To the contrary, when the undoped layer number is less than the doped layer number, the doped layer dominates the system and thus the $C=1$ phase appears.  Therefore, although the high-$C$ phases can be realized in these two kinds of the TI multilayer structures, their bulk gaps are rather small, which will hinder the experimental observations, as the Fermi energy needs to be tuned in the gap.  We suggest that for these two kinds of the TI multilayer structures, the system still lies in the gapless ``semimetal" phase.

\section{Conclusions}

To summarize, in this paper, we have developed an effective method to determine the Chern number and then obtained the reliable Chern number phase diagrams in the TI multilayer structures.  Compared with the commonly used Kubo formula~\cite{Thouless, Y.F.Zhao} or Fukui's algorithm~\cite{Fukui} to calculate the Chern number, the method of the Chern number determination developed in this work only requires to capture the evolutions of the phase boundaries.  Thus it needs less computational resources and is more efficient.  We believe that this method can be extended to other related topological systems.  The phase diagrams in Figs.~\ref{Fig3} and~\ref{Fig5} show that they have similar structures for different double-layer unit number $m$, which can provide a plethora of the high-$C$ phase transitions and need to be demonstrated in the future experiments.  

We admit that the interface-Dirac-states mechanisms were successful in explaining the high-$C$ phase modulation of the $m=2$ structure in the experiment~\cite{Y.F.Zhao}, but they cannot be deemed as the general mechanism for the high-$C$ behavior in the TI multilayer structure.  
Instead, the high-$C$ phase should be explained by the band inversions, which are similar to that in the TI thin film~\cite{J.Wang}.  We find that the high-$C$ behavior can also exist in two other kinds of the TI multilayer structures, but are difficult to be realized in the experiment due to their rather small bulk gaps.  Although the parameters in Bi$_2$Se$_3$ are used in the calculations, similar conclusions can be also expected for other TI material parameters, such as in Bi$_2$Te$_3$~\cite{C.X.Liu}.  There are also some open questions about the TI multilayer structures that remain to be resolved, such as the electron behaviors to the magnetic field and disorder.

\section{Acknowledgments} 

This work was supported by the National Natural Science Foundation of China (Grants No. 11804122 and No. 11905054), the Fundamental Research Funds for the Central Universities of China, and the China Postdoctoral Science Foundation (Grant No. 2021M690970).

 \begin{figure*}
 	\centering	
 	\includegraphics[width=18.4cm]{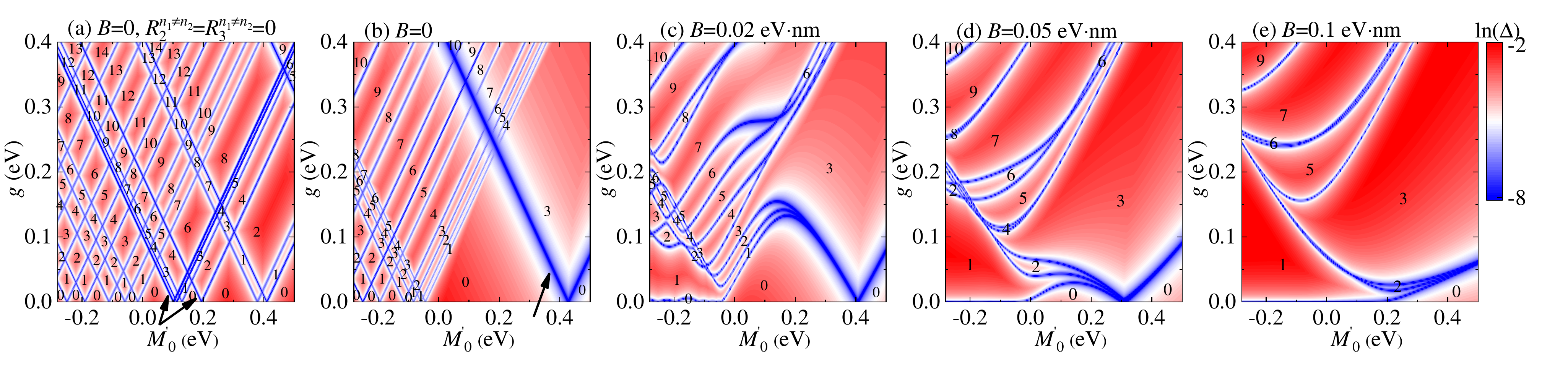}
 	\caption{Chern number determination of Fig. 3(c) in the main text, with the Chern number values being labeled in each figure.  From (a)$-$(e), the phase boundary evolutions are clearly observed, where (a) and (b) correspond to steps (i) and (ii), respectively, while (c)$-$(e) correspond to step (iii).  The contour scale represents the magnitude of ln$(\Delta)$, where $\Delta$ is the energy gap of the lowest bands.  The bright blue lines denote that the gap is closed and thus separate the different Chern number phases.  Note that when the phase boundary as labeled by the arrow(s) in (a) [(b)] is crossed, two (three) degenerated bands around the $\Gamma$ point are inverted and thus the Chern number will be changed by two (three).}
 	\label{Fig8}
 \end{figure*}

\section*{Appendix A: The explicit forms of $f_1$ and $f_2$}

The functions of $f_1$ and $f_2$ are defined as
\begin{align}
f_1^{n_1,n_2}=\sum_{i=1}^{m+1}
\int_{-\frac{L}{2}+(i-1)(d_1+d_2)}^{-\frac{L}{2}+id_1+(i-1)d_2}dz
\varphi^*_{n_1}(z) \varphi_{n_2}(z),
\end{align}
and 
\begin{align}
f_2^{n_1,n_2}=\sum_{i=1}^m
\int_{-\frac{L}{2}+id_1+(i-1)d_2}^{-\frac{L}{2}+i(d_1+d_2)}dz
\varphi^*_{n_1}(z) \varphi_{n_2}(z),
\end{align} 
with $\varphi_n(z)=\sqrt{\frac{2}{L}}
\text{sin}(\frac{n\pi z}{L}+\frac{n\pi}{2})$.  After straightforward calculations, we can obtain the explicit forms of $f_1$ and $f_2$. 

When $n_1=n_2$, we have
\begin{align}
f_1^{n_1,n_2}=&\frac{(m+1)d_1}{L}+\frac{1}{n_1\pi}
\text{sin}\big(\frac{d_2n_1\pi}{L}\big)
\nonumber\\
&\times\sum_{i=1}^m\text{cos}\Big[
\frac{(2id_1+2id_2-d_2)n_1\pi}{L}
\Big], 
\end{align}
and 
\begin{align}
f_2^{n_1,n_2}=&\frac{md_2}{L}-\frac{1}{n_1\pi}
\text{sin}\big(\frac{d_2n_1\pi}{L}\big)
\nonumber\\
&\times\sum_{i=1}^m\text{cos}\Big[
\frac{(2id_1+2id_2-d_2)n_1\pi}{L}
\Big], 
\end{align}
and when $n_1\neq n_2$, we have
\begin{align}
&f_1^{n_1,n_2}=
\frac{2}{(n_1^2-n_2^2)\pi}\sum_{i=1}^m\Big[
\nonumber\\
&
n_2\text{cos}\big(\frac{(id_1+id_2-d_2) n_2\pi}{L}\big)\text{sin}\big(\frac{(id_1+id_2-d_2) n_1\pi}{L}\big)
\nonumber\\
&-n_2\text{cos}\big(\frac{(id_1+id_2) n_2\pi}{L}\big)
\text{sin}\big(\frac{(id_1+id_2) n_1\pi}{L}\big)
\nonumber\\
&-n_1\text{cos}\big(\frac{(id_1+id_2-d_2) n_1\pi}{L}\big)
\text{sin}\big(\frac{(id_1+id_2-d_2) n_2\pi}{L}\big)
\nonumber\\
&+n_1\text{cos}\big(\frac{(id_1+id_2) n_1\pi}{L}\big)
\text{sin}\big(\frac{(id_1+id_2) n_2\pi}{L}\big)
\Big], 
\end{align}
and 
\begin{align}
&f_2^{n_1,n_2}=-f_1^{n_1,n_2}. 
\end{align}

\section*{Appendix B: An example of the Chern number determination}

As mentioned in the main text, three steps (i)$-$(iii) are adopted in determining the Chern number of a TI multilayer structure, with the main idea being to capture the evolutions of the phase boundaries.  In Fig.~\ref{Fig8}, as an example, we give more details for the Chern number determination of Fig.~\ref{Fig3}(c) in the main text.

In Fig.~\ref{Fig8}(a), when $B=0$ and the off-diagonal terms of $R_2$ and $R_3$ are absent, the phase boundaries are exhibited as the straight lines, which are consistent with the analytic phase boundaries of Eq.~(\ref{phaseboundary}) in the main text.  When there is no Zeeman splitting, $g=0$, the time-reversal symmetry is preserved and the system lies in the trivial $C=0$ phase.  With the increasing of $g$, if the Dirac mass $m_{n+}$ ($m_{n-}$) of an additional $n+$ ($n-$) subband is inverted, the Chern number will be increased (decreased) by one.  Note that the inverted bands crossing the phase boundaries labeled by the arrows are twofold degenerate.  Exactly, the phase boundary labeled by the left arrow is related to the accidental degeneracy of the $9+$ and $10+$ subbands around the $\Gamma$ point, while that labeled by the right arrow is related to the accidental degeneracy of the $5+$ and $6+$ subbands around the $\Gamma$ point.  Cases are the same for $n-$ subbands. 

In Fig.~\ref{Fig8}(b), when the off-diagonal terms of $R_2$ and $R_3$ are introduced, the linear phase boundaries are kept, but their positions are shifted.  When crossing the phase boundary labeled by the arrow, the three-fold degenerated bands around the $\Gamma$ point are inverted, thus the Chern number will be changed by three, from $C=0$ to $C=3$. 

In Figs.~\ref{Fig8}(c)$-$(e), when the coupling increases from $B=0$ to its realistic value $B=0.226$ eV$\cdot$nm, several aspects are worthy pointing out: (i) The linear phase boundaries are broken and the regions of the same Chern number are gradually connected. (ii) The high-$C$ phases are pushed upwards to enter the stronger $g$ regime.  To the contrary, the $C=1$ phase is pushed downwards to be close to the $g=0$ line, which shrinks the region spanned by the $C=0$ phase, meaning that a tiny magnetic-doping will drive the TI multilayer system to enter the $C=1$ phase.  (iii) The regions of $C=2$, $C=4$ and $C=6$ are shrinking and finally disappear, meaning that the inverted bands become twofold degenerate around the $\Gamma$ point when crossing the related phase boundaries.

\end{document}